\documentclass [twocolumn,showpacs,prl]{revtex4}
\usepackage{graphicx}

\begin{document}
\title{Laser Generated Magnetic Pulses: Comparison of Hot Electron Propagation in Conducting and Dielectric Material}   
\author{A. S. Sandhu,A. K. Dharmadhikari, G. R. Kumar}
\address{Tata Institute of Fundamental Research, 1 Homi Bhabha Road, Mumbai 400 005, India,}
\author{S. Sengupta, A. Das, and P. K. Kaw}
\address{Institute for Plasma Research, Bhat, Gandhinagar, Ahmedabad 382428, India}
\date{\today}

\begin{abstract}
	We report experimental evidence of electrostatic inhibition of fast 
electrons, generated in a highly resistive material upon irradiation with an 
intense ultra-short ($10^{16}\,W/cm^{2}$, $100\, fmsec$) laser pulse. The 
experiment involves measurement of temporal evolution of self-generated 
magnetic pulses using pump-probe polarimetry. 
A comparison is made between the temporal behaviour of magnetic pulses 
generated with Aluminum and Glass targets. It is found that in contrast to 
Aluminium, self-generated magnetic pulse decays much faster in glass. 
This is attributed to the absence of return shielding currents in glass, 
which results in build up of electrostatic field, which in turn inhibits the 
movement of fast electrons.	
Fitting of experimental measurements using a one dimensional model, yields 
estimate of conductivity of Aluminium and glass, and penetration depth of hot 
electrons in these materials. 
\end{abstract}
\maketitle

Interaction of intense ultrashort laser pulses with a solid leads to 
generation of large magnetic fields.  Magnetic fields up to 
gigagauss magnitudes have been 
predicted in the overdense region of solid density plasmas 
\cite{Sudan,Stamper}, which are created by explosive ionzation of a solid
target by an incident laser. Fields comparable to stellar magnitudes 
($\ge 300$ MG) have already been experimentally realized in the laboratory with 
the use of super-intense ultra-short laser pulses \cite{Tatarakis}. 
Recent studies 
have also established the ultrashort duration ($\sim picosec.$) of these 
self-generated magnetic 
fields under femtosecond laser excitation \cite{Sandhu,Borghesi}. The temporal 
behaviour of the magnetic pulse contains crucial information about the 
dynamics of hot
electron propagation inside a solid density material / plasma. This
information about the penetration and transport of hot electrons \cite{Mason} 
inside an
overdense plasma and factors affecting it, is of utmost importance to the
fast ignitor approach of fusion research \cite{Kolka,Tabak}. 

The pulsed high magnetic fields generated primarily near the critical 
density surface, result from a combination of currents, {\it viz.} the direct
hot electron current which is due to copious generation of fast electrons at
the critical layer via collisionless absorption of incident laser energy
, and the return shielding current which is generated in response
to the hot electron current \cite{Sentoku}. The propagation of hot electrons 
inside the 
solid density target depends on whether the target is a dielectric or a 
conductor. In case of a dielectric ( material with high classical resistivity ),
the return shielding current is weak, resulting in a large charge imbalance
which in turn exerts a strong retarding field on the fast electrons, resulting
in electrostatic inhibition.
Electrostatic inhibition of hot electrons in dense matter has been seen in some 
experiments  by measuring K $\alpha$ emission from layered targets. 
In initial experiments by Bond et. al. \cite{Bond} inhibition of fast 
electrons due to resistive electric field was observed in low density gold 
targets. In contrast to a dielectric, in a conductor ( material with extremely
low classical resistivity ), the return shielding current nearly balances the
hot electron current. Here the stopping of hot electrons occurs because of
turbulence induced anomalous resistivity, the chaotic magnetic turbulence
effects being generated by the presence of large gradients in the plasma 
currents \cite{Drake,Sentoku2}. 
Experimental evidence of turbulence induced
anomalous resistivity is presented in a recently published experiment by
Sandhu et. al. \cite{Sandhu}, where they irradiate a solid aluminium target
with a high intensity femtosecond laser. Since the mechanism of hot electron
transport ( along with return shielding current ) and their eventual stopping 
is different ( it is either electrostatically induced \cite{Bond} or
turbulence induced \cite{Sandhu} ) in a dielctric and a conductor, it is 
expected that the temporal evolution of the magnetic fields which these
currents generate will also be different. Thus the complex dynamics of the hot
electron propagation inside a solid density material ( or plasma ) is reflected
in the time evolution of the magnetic pulse.

Here we present experimental results on transport of hot electrons and their 
inhibition in both high and low resistivity media via measurements on temporal
evolution of megagauss magnetic pulses. Recently, there has been a lot of 
experimental work on the transport of hot electrons \cite{Pisani}, but to our
knowledge no experimental result exists which gives a clear distinction 
between electrostatically induced and turbulence induced inhibition of fast
electrons. In our experiment, this has become possible through the use of
temporal measurement of magnetic pulses which is intimately related to the
propagation of hot electrons inside the material.
We compare two entirely different media, one conducting (solid Aluminum) and 
the other insulating (BK7 Glass). We use pump-probe polarimetry to measure the 
ellipticity produced in a probe laser pulse due to pump-generated magnetic 
fields. This measured ellipticity is then analyzed numerically to deduce the 
magnetic field as a function of time delay between pump and probe pulses. 
We find that the evolution of magnetic field in aluminium and glass to be 
drastically different. The time required for the magnetic field to decay in 
glass is found to be an order of magnitude smaller than in aluminium. This is 
a clear indication of electrostatic inhibition of fast electrons which results
from absence of return shielding currents due to very low electrical 
conductivity of glass.  We also present an 
one-dimensional model ( an improvement over our zero-dimensional model 
\cite{Sandhu} )  for magnetic field evolution which gives an interpretation of 
our observations. The decay time in each case is related to the electrical 
conductivity ( electrostatically induced or turbulence induced ), thus 
providing us with estimates of conductivity of glass and aluminium under 
conditions of elevated temperatures and pressures. 

\begin{figure}
\includegraphics [width=2.8in,height=3in]{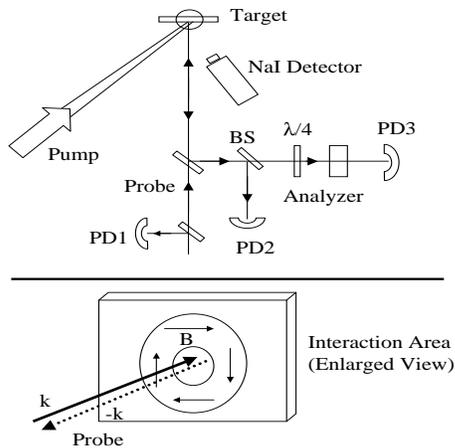}
\caption{Experimental setup, Target (T), Lens (L), Beam Splitter (BS), Photo Diode (PD), Half Wave Plate (HWP). Inset represents pump and probe paths in plasma. Expanded view of interaction region shows magnetic field geometry.}
\end{figure}

The experimental set-up is shown in figure 1. The p-polarized pump laser 
(100fs, 800nm) is focused at an angle of $45^{\circ}$, to an intensity of 
$\sim 2 \times 10^{16} Wcm^{-2}$ on a spot size of 20 microns. The long term 
contrast for the pump is $10^{5}$. The probe laser used in measurements is 
frequency doubled (400nm) in a thin BBO crystal, and is made $\sim 10^3 - 10^4$
times weaker than the pump intensity and is focused at near normal incidence. 
In this configuration, the probe penetrates beyond the critical density 
($n_{c}$) for the pump laser (i.e. upto 4$n^{(806)}_{c}$), and thus samples 
high density region near and beyond the critical layer, where high magnetic 
fields are expected to exist. It should be noted that this high density region 
is not accessible using tangential polarimetric measurements of Faraday 
rotation. The input laser fluctuations are monitored with a photodiode (PD1). 
The reflected probe from the interaction region is split into two arms, one 
arm is used in PD2 to measure the reflectivity of the probe and the 
second arm is 
used in the quarter wave plate and polarizer combination ( kept in front of 
photodiode PD3 ) to measure the polarization parameters of the probe. 
The reflectivity measurement serves to establish the zero delay and is used to 
normalize other polarimetric measurements. All the measurements are made as a 
function of time delay between pump and probe. Diffuse scatter is also 
monitored with another PD and is observed to be insignificant and constant in 
our temporal range of interest. In our experiment, we have used two type of 
targets, first one an aluminium target i.e. a conductor and second one a glass
target i.e. an insulator. The BK7 glass target was coated with about half a 
micron thick 
aluminium on the front surface so that laser hitting the target generates 
identical hot electron temperatures in both cases. To establish that this is 
indeed
so, we have measured the hard X-rays arising from hot electrons in the energy
range from 50 to 500 keV for both the 
cases, under similar laser conditions with a NaI(Tl) detector. 
The hot electron spectra are  shown in figure 2(a). The hot electron temperature 
obtained from the  fit (shown as solid line) is observed to be approximately 
same ($\sim 30 keV$) in both the cases (Al layer on Glass and Al metal alone). 
This indicates that we 
have identical hot electron source in both cases arising from interaction of 
the laser pulse with the front surface of the target. 
The magnetic field induced ellipticity in the probe pulse is measured using 
different combinations of quarter wave plate and polarizer in analyzer setup i.e. polarizer at $ 0, 45, 90^\circ$ and $\lambda /4$ plate at $45^\circ$ and 
polarizer at $90^\circ$ \cite{Segre}. The measurement of stokes parameters in 
this method rules out the presence of any random depolarization.  
Experimental curve showing ellipticity vs time delay is also shown in figure 2(b). 
The significant difference is observed in time scale when we compare the two cases, that of glass and aluminum.
We now numerically analyze the measured ellipticity in order to deduce the 
magnetic field present in the interaction region. 

\begin{figure}
\includegraphics [width=2.8in,height=3in]{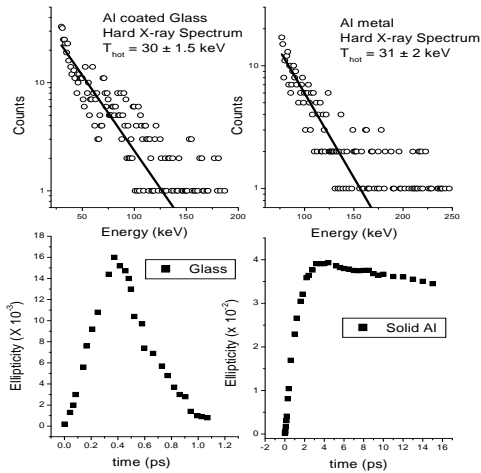}
\caption{The hard X-ray emmission spectra (a) for glass sample coated with 0.6 micron Al (b) pure Al metal target; in energy range 50-500 keV with laser intensity of $2 \times 10^{16}$ Wcm$^{-2}$.  Solid line is the exponential fit to deduce the hot electron temperature in each case. (c) Raw data for ellipticity in case of Glass coated Al (d) in case of solid Al metal. }
\end{figure}

An electromagnetic wave travelling through a magnetized plasma with wave
vector perpendicular to the quasistatic magnetic field can be analyzed in terms
of two characteristic modes, the ordinary mode (O-wave) with electric field 
polarization
vector parallel to the magnetic field and the extraordinary mode (X-wave) with 
electric field polarization vector perpendicular to the magnetic field 
\cite{Chen}. These two modes have different refractive indices, which  
are
\begin{equation}
\mu_{O} = \left[1 - \frac{\omega_{pe}^{2}}{\omega^{2}}\right]^{1/2} 
\end{equation}
\begin{equation}
\mu_{X} = \left[1 - \frac{\omega_{pe}^{2}(\omega^{2}-\omega_{pe}^{2})}{\omega^{2}(\omega^{2}- \omega_{pe}^{2}- \omega_{ce}^{2})}\right]^{1/2}
\end{equation}
where $\mu_{O}$ represents refractive index for $O$-mode and $\mu_{X}$ 
represents refractive index for $X$-mode. Here $\omega$, $\omega_{pe}$ and
$\omega_{ce}$ are respectively the laser frequency, the electron plasma 
frequency, and the electron cyclotron frequency i.e. $eB/mc$.
The difference in refractive indices results in accumulation of phase 
difference between O- and X- waves as they propagate through the plasma. This 
difference in phase is observed as induced ellipticity. If we represent the 
laser polarization in terms of stokes vector $s = (s_{1}, s_{2}, s_{3})$ , 
$s_{1} = cos(2\chi) cos (2\psi)$, $s_{2} = cos(2\chi) sin (2\psi)$ , 
$s_{3} = sin(2\chi)$, where $\psi$ is orientation of polarization ellipse, 
then the ellipticity $\epsilon$ is given by $\epsilon = tan(\chi)$. The 
evolution of stokes vector inside the magnetized plasma is given by the 
difference in refractive indices of the characteristic waves 
i.e. $ ds/dz = \Omega(z) \times s(z)$, where $\Omega = (\omega /c)(\mu_{O} 
-\mu_{X})$. The solution of this polarization evolution equation is written as 
$s(z) = M . s_{input}$, where $M_{ij}$ is the transition matrix which obeys 
the equation $ dM_{i}/dz = \Omega \times M_{i}$, where $M_{i} = (M_{1i} , 
M_{2i} , M_{3i})$ \cite{Segre}. We integrate this equation numerically inside 
the plasma by dividing it into small slabs, where within each slab, the plasma 
parameters are assumed to be constant.

For the purpose of integration, the required plasma density profile is 
modelled by 
assuming that the plasma expands into vacuum at ion sound speed in a self
similar fashion \cite{Mora,Kruer}, which yields an exponential density 
profile of the form $ n_{e} = n_{s} exp(-z/c_{s}t)$, where $n_{s}$ is the solid 
density and $c_s = \sqrt{(Z kT_{e}/m_{i})}$ is the ion sound speed. We have 
chosen the background electron temperature $T_{e} = 100 eV$ \cite{Multi} and 
the average 
ionization state $Z=5$ \cite{Milchberg}. At the vacuum end, which is chosen at 
a distance $L_{s}$ from the critical layer, we put a cut off 
on the density `$n_{end}$' which is chosen sufficiently low so that it makes
practically no difference to the results.
The plasma slab to be integrated over is given by 
$L_{s} = c_{s} t [ln(n_{critical}/n_{end})]$. It is found that the results are 
not sensitive to the chosen density profile; even the use of linear density
profile yields results within acceptable error bars. In addition to the density
profile, we have assumed uniform magnetic field inside the plasma column. 
Although, this assumption is not strictly correct, it makes little difference 
to the results,  and is a reasonable one as shown later. The solution  
$M_{ij}(z)$,
and $s(z) = M . s_{input}$ can now be determined for any small plasma slab (
where the density is assumed to be constant ) for a given value of magnetic 
field B in definition of $\Omega$. Thus recursively solving in small steps and 
maintaining unitarity of M one can find the full transition matrix for whole 
plasma column, for any chosen density profile. From the transition matrix, the 
final stokes vector can be computed which in turn gives the ellipticity of the 
reflected probe.

In addition to the phase difference introduced by differences in refractive 
indices, there is an extra phase difference which aries due to slightly 
different turning points for O and X-wave. The X-mode has a turning point i.e. 
right cut-off in the dispersion diagram, earlier than the O-mode, 
which reflects 
at usual critical density. This extra path travelled by O-mode as compared to
X-mode leads to an additional phase difference, which we have included in our
calculation of net ellipticity. Thus the effect of magnetic field manifests 
itself in the form of ellipticity of the reflected probe light by both of the 
above mentioned processes i.e. difference in refractive indices and difference 
in turning points.
%
%

Using the above described procedure, for each time delay, we iteratively 
calculate the magnetic field B required to yield the experimentally observed 
ellipticity. The final plots of magnetic field for solid aluminium target and 
BK7 glass coated with aluminium layer are shown in fig. 3(a) and fig. 3(b) 
respectively. The squares represent experimental points
and the solid curves, which show a reasonable fit to the experimental data
are obtained using a one-dimensional model developed below. It is observed that
the peak magnetic field reached in both cases are of the same order ( $32 MG$ 
in case of BK7 glass and $22 MG$ in case of solid aluminium ). The most 
striking difference between the two cases is seen in the time durations of the
self-generated magnetic pulses (note different time axis of fig 4(a) and (b)). 
In case of BK7 glass coated with alumunium, the time duration of the magnetic
pulse ($\sim 1 picosec$) is an order of magnitude smaller than the time 
duration ($\sim 10 picosec$) of the magnetic pulse for pure aluminium. This is 
a clear indication of high background resistivity of glass, which curtails the 
return currents and hence results in inhibition of hot electrons because of 
electrostatic fields. This leads to much faster resistivite decay of magnetic 
fields in case of glass sample as compared to solid aluminium sample, where the
decay is mainly due to turbulence induced anomalous resistivity \cite{Sandhu}.
\begin{figure}
\includegraphics [width=2.8in,height=3in]{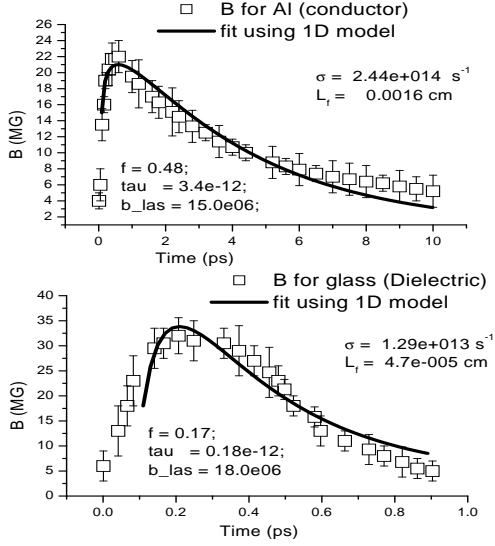}
\caption{Magnetic field pulse profile for (a) Glass (b) Aluminum with laser with intensity of $2 \times 10^{16}$ Wcm$^{-2}$. Solid line shows the fit for p-polarized case using a phenomenological model.}
\end{figure}

We now model the temporal evolution of magnetic field using the following 
equation, which describes the mechanism of quasistatic magnetic field
generation under EMHD approximation.  
\begin{equation}
\frac{\partial \vec{B}}{\partial t} = \frac{c^{2}}{4 \pi \sigma} \nabla^{2} 
\vec{B} + \frac{c}{\sigma}\left( \vec{\nabla} \times \vec{j}_{hot} \right) 
\label{mag1}
\end{equation}
The above equation is derived by taking the curl of the equation of motion
of background plasma electrons that carry the return shielding currents,
$\vec{j}_{p} = (c/4 \pi)(\nabla \times \vec{B}) - \vec{j}_{hot}$, where
$\vec{j}_{p} = \sigma \vec{E}$, $\sigma$ being the conductivity of the 
background plasma. The first term describes the magnetic field decay due to
resistive damping of the plasma shielding currents and the second term 
describes the magnetic field generation due to hot electron currents. Assuming
the magnetic field to be in the azimuthal direction, equation (\ref{mag1}) can
be approximated as
\begin{equation}
\frac{\partial B}{\partial t} \approx - \frac{B_{\phi}}{\tau} + S(z,t)
\label{mag2}
\end{equation}
where the diffusion term is approximated as $B/\tau$ with $\tau 
\approx (4 \pi \sigma / c^{2})(\Delta r)^{2}$, and the source term is 
approximated as $S(z,t) \approx - (c / \sigma (\Delta r)) j_{hot}(z,t)$. Here
$\Delta r$ is the laser spot radius $\approx 10 \mu$. Our main interest lies
in understanding the evolution of the magnetic field after the laser is 
switched off. This is because the experimental results show that, 
although the laser is on only for $\sim 100 fmsec.$, the magnetic field peaks
and remains on for picoseconds time duration.  
Taking $B=B_{las}$ at $t=\tau_{laser}$, the solution of equation (\ref{mag2})
is given by
\begin{eqnarray}
B  & = & B_{las} e^{-(t - \tau_{laser})/\tau} + e^{-t/\tau}\int^{t}_{\tau_
{laser}} S(t,z,r) e^{t/\tau} dt \nonumber \\
& & \nonumber \\
& \approx & B_{las} e^{-(t - \tau_{laser})/\tau} - e^{-t/\tau}\int^{t}_{\tau_
{laser}} \frac{c}{\sigma \Delta r} j_{hot} (z,t) e^{t/\tau} dt  \label{mag3}
\end{eqnarray}
where $j_{hot} = - e n_{h} v_{h}$, $n_{h}$ is the hot electron density and
$v_{h}$ is the velocity of the hot electron fluid.
To make an estimate of $n_{h}$ and $v_{h}$, we use the formalism given by 
Bell et. al. \cite{Bell}. According to \cite{Bell}, the evolution of 
hot electron density ($n_{h}$) is governed by the following nonlinear 
diffusion equation
\begin{equation}
\frac{\partial n_{h}}{\partial t} = \frac{\partial}{\partial z} \left(
\frac{\sigma T_{h}}{e^{2} n_{h}} \frac{\partial n_{h}}{\partial z} \right)
\label{mag4}
\end{equation}
Here $T_{h}$ is the hot electron temperature. 
It can be shown that the above equation is valid even without the restricted
assumption $j_{total} = j_{hot} + j_{p} \approx 0$, used in ref. \cite{Bell}.
Since our interest lies in $t >  \tau_{laser}$, we use the solution of 
equation (\ref{mag4}) in this temporal regime, as given in ref. \cite{Bell}
\begin{equation}
n_{h} = \frac{2 n_{0} z_{0}}{\pi}\frac{L}{z^{2} + L^{2}}
\label{mag5}
\end{equation}
with
\begin{eqnarray}
L(t) & = & z_{0} \left[ \frac{5 \pi \sigma T_{h}}{3 e^{2} n_{0} z_{0}^{2}}
\left(t - \tau_{laser}\right) + 1 \right]^{3/5}\\
& & \nonumber \\
n_{0} & = & \frac{2}{9} \frac{I_{abs}^{2} \tau_{laser} e^{2}}
{\sigma T_{h}^{3}} \\
& & \nonumber \\
z_{0} & = & \frac{3 \sigma T_{h}^{2}}{e^{2} I_{abs}}
\end{eqnarray}
where the absorbed intensity $I_{abs} = f I_{incident}$, $f$ being the 
fraction absorbed. Here $n_{0}$ is the density of hot electrons at
$z=0$, at time $t = \tau_{laser}$ and $z_{0}$ is the characteristic stopping 
length such that $n_{0}z_{0}$ is the total number of hot electrons 
produced at time 
$t = \tau_{laser}$. The constants $n_{0}$ and $z_{0}$ have been derived by 
equating the absorbed laser energy ``$I_{abs} \tau_{laser}$'' to the hot
electron kinetic energy (see ref. \cite{Bell}). The above solution (\ref{mag5})
is a self-similar solution of equation (\ref{mag4}) in which the spatial shape
remains the same but it expands in time with a scale length $L(t)$.
Using the above expression for $n_{h}$ and $L(t)$, we estimate $j_{hot}$ as
\begin{eqnarray}
j_{hot} & = & - e n_{h} v_{h} \nonumber \\
& & \nonumber \\
& = &  - e \frac{2 n_{0} z_{0}}{\pi} \frac{L}{z^{2} 
+ L^{2}} \left[ \alpha \frac{d L}{d t}\right]
\end{eqnarray}
where $v_{h}$, the hot electron velocity is taken to be proportional to 
$\frac{d L}{d t}$, $\alpha$ being the proportionality constant. Substituting 
the expression for $\frac{dL}{dt}$ in $j_{hot}$ and using it in equation 
(\ref{mag3}) along with  $\sigma \approx \frac{c^{2}}{4 \pi (\Delta r)^{2}} 
\tau$, we get
\begin{equation}
B = B_{las} e^{-y} + A e^{-y} \tau \int^{y}_{0} \frac{(p y \tau + 1)^{1/5}
e^{y}}{z^{2} + z_{0}^{2}(p y \tau + 1)^{6/5}} dy
\label{mag6}
\end{equation}
where $ y  =  \frac{(t - \tau_{laser})}{\tau} $, $ A =  \frac{2 c z_{0} 
\alpha T_{h}}{e \Delta r} $ and $ p = \frac{5 \pi \sigma T_{h}}{3 e^{2} n_{0} 
z_{0}^{2}} $.
We now use the above expression for $B(t,z)$ at $z = 0$ to model the 
magnetic field
evolution as a function of time, for both BK7 glass coated with aluminium and 
solid aluminium using $\tau$ ( which is related to conductivity $\sigma$ )
and $f$ ( fraction of light absorbed ) as free parameters. 
$z=0$ is the point where hot electrons are generated; so it is actually the 
critical density point. The proportionality contstant $\alpha$ is taken to be 
unity.
$B_{las}$, the magnetic field at time $t = \tau_{laser}$ is $15 MG$ and 
$18 MG$ respectively for aluminium and glass. The best fit curves ( solid 
lines ) are shown in fig. 3(a) and fig. 3(b) respectively. From the best fit, 
the relevant parameters ( $f$ and $\tau$ ) for aluminium and glass are:  

Aluminum:  
f = 0.5; $\tau$  = 3.5 ps

Glass:   
f = 0.2; $\tau$ = 0.2 ps  

Before presenting a discussion of our results obtained from one-dimensional
modelling and numerical fit, we first justify the assumption we have made in
our numerical analysis of ellipticity {\it viz.} the assumption of uniform
magnetic field inside the plasma slab. Fig. 4 shows the typical magnetic field
as a function of $z$ and $t$ obtained from our one-dimensional model. This
plot is for aluminium with $f$ and $\tau$ as calculated above. As expected,
the magnetic field diffuses in $z$ and decreases in amplitude as time increases.
Although the magnetic field $B$ varies with $z$, we shall see below that its
variation is not significant over the range of scale lengths which are relevant
for our experiments. In Fig. 5, using equation (\ref{mag6}) we have plotted 
the magnetic field as a function of position at a time corresponding to the
peak of the magnetic field, for both aluminium and glass. In this plot, we have
used the same value of $f$ and $\tau$ as obtained above. This plot clearly
shows that the variation of magnetic field over the plasma slab is negligible.
This justifies that the assumption of uniform magnetic field in our analysis of 
ellipticity is a reasonable one.

\begin{figure}
\includegraphics [width=2.8in,height=3in]{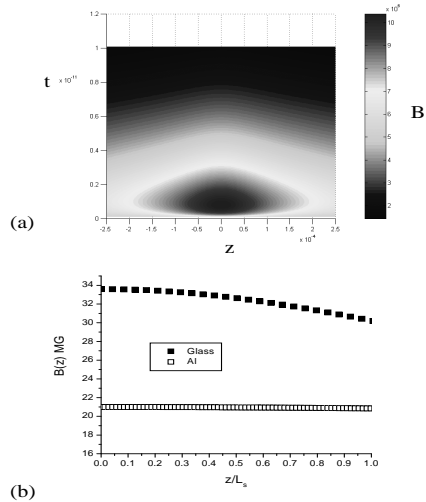}
\caption{(a) The z and t dependence of the magnetic field profile inside the plasma based on one dimensional model for propagation of hot electrons as disscused in the text.(b) Spatial dependence at the time corresponding to the peak magnetic field in both the cases.}
\end{figure}

We now discuss our results. From the results of the fit, it is clear that the
time for magnetic field decay in aluminium is an order of magnitude larger than
in glass. Calculation of conductivity using $\sigma = ( c^{2} / 4 \pi 
(\Delta r)^{2} ) \tau$, gives $\sigma_{Al} = 2.5 \times 10^{14} s^{-1}$ and
$\sigma_{glass} = 1.43 \times 10^{13} s^{-1}$.
As of now there are no measurements of conductivity at elevated temperatures 
for glass and very few in aluminium. We see that resistivity of aluminium
as deduced from these measurements is an order of magnitude higher than the
value of resistivity deduced from reflectivity measurements by Milchberg 
et. al. \cite{Milchberg}. This, we believe, is due to turbulence induced 
anomalous resistivity which affects the return currents. This conclusion is 
also supported by simulation where similar effect of anamalous 
resistivity and stopping of hot electrons have been seen \cite{Sentoku}. 
In the case of glass, the situation is more complex as at low temperatures 
glass is almost non-conducting $(\sigma \sim 0)$. However, at elevated 
temperatures, as a result of target heating by collisional effects and large 
electric fields exceeding breakdown threshold, conductivity becomes finite. 
There exists observation of $8 \mu$ wide collimated ionization tracks 
extending upto $150-300 \mu$ and  lasting for $60\, ps$ in glass 
with a $10^{17} W cm^{-2}$ intensity laser \cite{Teng}. 
Although we cannot estimate here the amount of target heating or ionization, 
we do get an estimate of resitivity, which we assume to be uniform in space 
and time. Our measurements show that neutralization of hot electron current is 
clearly not as effective in glass as it is in aluminium, due to high 
background resistivity. This results in inhibition of hot electron propagation 
in glass through generation large of electrostatic fields. 
We now make an estimate of the penetration depth ($L_{f}$) of hot electrons 
using equation (8). Using the parameters $z_{0}$ and $\sigma$ from the fit and 
$T_{hot}$ from observations, we get, for glass, $L_{f} = 4.7 \times 10^{-5} cm$ 
in $1\, ps$ and $2.1 \times 10^{-4} cm$ in $10 ps$, and for aluminium 
$L_{f}= 1.6 \times 10^{-3} cm$ in $10 ps$. These numbers indicate that hot
electrons are penetrating an order of magnitude more distance in a conducting 
background. 
Similar results on electrostatic inhibition have also been measured in plastic 
targets using $ K_{\alpha}$ emission from layered targets. This technique  
has its limitations due to its reliance on x-ray measurements which is a 
secondary diagnostic (with limitations of sensitivity, calibration, and 
reabsorption) and on numerical codes for interpreting X-ray data. 

%
In conclusion, the penetration of fast electrons into a target can 
result in very high electric field, which is not quenched by thermal return 
currents. This can lead to severe impediment in the flow of fast electron 
currents. We provide evidence of inhibition of fast electrons in an 
insulating media. Our results on magnetic pulse measurement and modelling in 
terms of hot electron currents as source, yield measurements of conductivity 
of conducting and dielectric media under extreme conditions.  
These results are very important for laser fusion schemes, which rely on 
ignition of hot spot by energy deposition by fast electrons.

We would like to thank D. Mathur, M. Krishnamurthy,and  P. P. Rajeev for 
discussions and useful suggestions.

\end{document}